\documentclass{kluwer}    

\newdisplay{guess}{Conjecture}

\begin{document}                                                                                   
\begin{article}
\begin{opening}         
\title{The Multiwavelength Approach to Unidentified Gamma-Ray Sources} 
\author{David J. \surname{Thompson}}  
\runningauthor{D. J. Thompson}
\runningtitle{Multiwavelength Approach}
\institute{Laboratory for High Energy Astrophysics\\NASA Goddard Space Flight Center\\Greenbelt, Maryland 20771 U.S.A.}
\date{July 2, 2004}

\begin{abstract}
As the highest-energy photons, gamma rays have an inherent interest to astrophysicists and particle physicists studying high-energy, nonthermal processes.  Gamma-ray telescopes complement those at other wavelengths, especially radio, optical, and X-ray, providing the broad, mutiwavelength coverage that has become such a powerful aspect of modern astrophysics. Multiwavelength techniques of various types have been developed to help identify and explore unidentified gamma-ray sources.  This overview summarizes the ideas behind several of these methods. 
\end{abstract}

\end{opening}           

\section{Introduction:  Gamma-Ray Sources as Multiwavelength Objects}

In the MeV range and above, astrophysical sources are almost exclusively non-thermal, {\it i.e.} produced by interactions of energetic particles, through such processes as bremsstrahlung, synchrotron radiation, Compton scattering, and neutral pion production.  High-energy, nonthermal sources are inherently multiwavelength objects, for several reasons:

\begin{enumerate}
\item Nature rarely produces monoenergetic particle beams. The cosmic ray spectrum, for example, covers a huge expanse of energies. A broad range of particle energies naturally leads to a broad range of photon energies. 

\item Charged particles rarely interact by only one process.  Different processes radiate in different energy bands. Blazars, for example, shine by synchotron radiation (electrons in magnetic fields) for radio through soft X-ray energies, while the same electrons Compton scatter low-energy photons to produce hard X-ray and gamma-ray emission. 

\item High-energy particles, as they lose energy, can radiate in lower-energy bands. The afterglows seen in gamma-ray bursts illustrate this effect.
\end{enumerate}

Results from the four instruments on the Compton Gamma Ray Observatory revealed a large number of sources, including gamma-ray bursts, pulsars, blazars, interstellar clouds, and a sizeable number that remained unidentified. In the third EGRET catalog, for example, over half the sources were considered unidentified \cite{Hartman}. Except for gamma-ray bursts, which are self-identifying, the other sources were identified by association with objects at other wavelengths.  The new knowledge of the Universe that came from exploring these sources usually involved combining the gamma-ray data with those from other parts of the electromagnetic spectrum. Multiwavelength studies are therefore both expected and essential for gamma-ray astrophysics. 

The following sections will describe some of the methods used for identification of unidentified gamma-ray sources.  In addition, some comments are included about how multiwavelength studies are used to move beyond identification to exploration of sources once they are identified. This paper is intended as a broad overview of the many facets of the multiwavelength approach, not an in-depth review.  Many of the latest results appear in other contributions to these proceedings. 

\section{X-Rays: Searches for Counterparts from the Highest Energies Down}

As the nearest neighbors to gamma rays in the electromignetic spectrum, X-rays are a logical starting point for gamma-ray source identification. The basic concept of starting with X-rays is that, at some level, gamma-ray sources will all have X-ray counterparts. If the X-ray counterpart can be found, the excellent X-ray position information allows deep searches at still longer wavelengths. 
The way this approach is applied is as follows:  using an X-ray image of a gamma-ray source error box, eliminate most of the X-ray sources from consideration based on their X-ray, optical, and radio properties.  Look for a nonthermal source with a plausible mechanism to produce gamma rays.
The classic example is Geminga. Bignami, Caraveo, and Lamb \shortcite{Bignami} started this search in 1983.  The final result appeared in 1992 with the detection of pulsations from this isolated neutron star in X-rays \cite{Halpern92} and then in gamma rays \cite{Bertsch}. 

A more recent example is the work on 3EG J1835+5918.  Parallel effort by two groups, \cite{Mirabal} and \cite{Reimer01}, used the same approach and reached the same conclusion.  Steps in the process were:

\begin{itemize}
\item A long ROSAT exposure was used to find candidate X-ray counterparts.
\item Optical observations were made of the X-ray counterparts.
\item All but one of the candidates were found to be unlikely gamma-ray emitting objects such as stars and QSOs.
\item The remaining candidate, RX J1836.2+5925, had no optical counterpart, suggesting an isolated neutron star.
\item A radio search of the error box revealed no radio pulsar counterpart \cite{Nice}.
\item A deep Chandra observation found a two-component spectrum, a thermal plus power-law combination.
\item The full multiwavelength Spectral Energy Distribution (SED) resembles that of Geminga.
\item Despite the absence of pulsations, this object appears to be an isolated neutron star, possibly a radio-quiet pulsar that will be seen in X-rays and gamma-rays with deeper exposures.
\end{itemize}
      
Other examples of the use of X-rays for ``top down'' gamma-ray source identification include: 
\begin{itemize}
\item ASCA observations of unidentified Galactic gamma-ray sources that show evidence of pulsar wind nebulae \cite{Roberts}.
\item Identification of a blazar near the Galactic Plane \cite{Mukherjee}.
\item XMM Newton observations of intermediate-latitude sources, with the analysis still in progress \cite{La Palombara}.
\end{itemize}

\section{Radio: Searches for Counterparts from the Lowest Energies Up}

The largest class of identified gamma-ray sources is blazars, all of which have radio emission characterized by a bright compact source and a flat radio spectrum.  
If a flat-spectrum radio source with strong ($\sim$1 Jy), compact emission at 5 GHz or above is found in a gamma-ray source error box, it becomes a blazar candidate. 
The approach is then to  use radio catalogs to search for flat-spectrum radio sources.  If a candidate is found, follow-up observations can be made to locate other blazar characteristics such as polarization, time variability and a Spectral Energy Distribution with a synchrotron plus a Compton component. 
The EGRET team used this approach in compiling the EGRET catalogs. Mattox, Hartman, and Reimer \shortcite{Mattox} quantified the method based on proximity and radio intensity. Sowards-Emmerd, Romani, and Michelson \shortcite{Sowards-Emmerd} have expanded the number of known blazars with this approach. 

One example of this type of identification is described by Foreman et al. \shortcite{Foreman} for 3EG JJ0433+2908:

\begin{itemize}  
\item From radio catalogs, a flat-spectrum source with 5GHz flux of 475 mJy was found.
\item Optical and X-ray catalogs showed a counterpart for this source.
\item Infrared observations also found a counterpart.
\item Follow-on radio and X-ray observations indicated likely variability.
\item Follow-on optical observations found a featureless spectrum.
\item The full multiwavelength Spectral Energy Distribution (SED) resembles that of blazars, with a synchrotron and a Compton component.
\item This appears to be a BL Lac object.
\end{itemize}
 
Radio observations have also been used to search ``bottom up'' for:

\begin{itemize}
\item  Extended gamma-ray sources related to molecular clouds \cite{Digel}
\item  Microquasars in gamma-ray source error boxes \cite{Paredes}
\item Possible supernova remnants with nearby molecular clouds \cite{Combi}; \cite{Torres}.
\end{itemize}

\section{Periodicity as a Multiwavelength Tool}

Finding a high-significance common periodic variability, either rotational or orbital, at multiple wavelengths is a definitive identifier for a source.  
Pulsars (rotating neutron stars) are the prototype of this method, dating back to balloon and early satellite observations of gamma rays from the Crab and Vela pulsars. 
The approach is to search out a periodic signal at one wavelength, then fold data from other wavelength bands at the expected period. For a summary of observations of gamma-ray pulsars, see \cite{Thompson}.
Thus far, all gamma-ray pulsars were first found at other wavelengths, where the higher density of photons made periodicity searches easier.  The GLAST Large Area Telescope (LAT) will have the capability for independent period searches for some gamma-ray sources. The deepest searches for pulsation will still require contemporaneous radio or X-ray observations.

Recent TeV observations with H.E.S.S. have indicated a source with orbital periodicity. The binary system consisting of PSR B1259$-$63 and a Be star shows gamma-ray emission near periastron \cite{Beilicke}. 

In addition to identification, periodic sources facilitate exploration of physical processes in the region of the neutron star. The light curves provide information about the physics and geometry of particle acceleration and interaction.  The broadband energy spectra show multiple components, produced by different physical processes.

\section{Flaring: Other Types of Varaibility}

The basic concept for using non-periodic variability is that transient or long-term variability helps to identify sources and to understand their physics. 
Variability studies have shown that gamma-ray pulsars have low variability (except for their pulsations), while blazars typically show substantial variability. 
The approach is to use degree of variability as a guide to source identification, then examine variability correlations at different wavelengths as a diagnostic of emission processes. 
Transients, including gamma-ray bursts, require fast response in order to obtain multiwavelength data. 

An example of using flaring as a multiwavelength tool to identify a blazar was given by Wallace et al \shortcite{Wallace}.  Their steps:
\begin{itemize}
\item In a study of short-term variability of EGRET sources, 3EG J2006$-$2321 was seen to be variable.
\item From radio catalogs, two flat-spectrum sources with modest 5GHz flux were found.
\item Optical observations showed one of these to be a normal galaxy (unlikely to be the gamma-ray source) and the other to have a redshift z = 0.83.
\item Observations of optical polarization from the distant object showed significant polarization, with variability.
\item X-ray observations showed only an upper limit.
\item The full multiwavelength Spectral Energy Distribution resembles that of blazars, with a synchrotron and a Compton component.
\item This appears to be a flat-spectrum radio quasar.
\end{itemize}

Blazars are characterized by both short-term and long-term variability at essentially all wavelengths.  The relationship between changes at different wavelengths is a powerful tool for studying the jets in these sources. This multiwavelength variability is an example of moving from identification to study of the detailed astrophysics of objects.

\section{Population Studies: Collective Properties of Sources}

Because gamma-ray source error boxes are typically large by standards of other branches of astronomy, positional agreement with a candidate object is rarely a strong argument by itself for physical association. 
Even if individual gamma-ray source identifications are not possible, however, statistical analysis may show a pattern of correlation with a given type of object. 
The approach is to  compare a gamma-ray source catalog (or subset of a catalog) with known classes of objects, using some statistical measure of correlation. 
If an association can be found, then it is valuable to study the implications of this source class. 

Some examples of population studies are:
\begin{itemize}
\item Kaaret and Cottam \shortcite{Kaaret} found a correlation between EGRET unidentified low-latitude sources and OB associations.
\item Yadigaroglu and Romani \shortcite{Yadigaroglu} also found evidence of gamma-ray sources near OB associations, as well as pulsars and supernova remnants.
\item Romero et al. \shortcite{Romero} also related EGRET sources to supernova remants and OB associations in a statistical sense.
\item At intermediate Galactic latitudes  Grenier \shortcite{Grenier} and  Gehrels et al. \shortcite{Gehrels} found that steady gamma-ray sources appear to represent a new population, distinct from the sources along the Galactic Plane. These sources may be associated with the Gould Belt. If so, they would be nearby, low luminosity objects, possibly pulsars or microquasars.  No individual candidate objects have yet been identified at other wavelengths. 
\end{itemize}

\section{Theory and Modeling: `Glue' That Connects the Various Observations}

Although multiwavelength observations are critical, theory and modeling are the means to link the different observations and construct a physical picture of a source.  Three broad areas show the importance of this work:

\begin{itemize}
\item Theoretical work provides explanation and understanding of what has already been seen. 
These calculations help move science from ``What is there?'' to ``How do they work?'' Gamma-ray pulsar modeling, for example, has evolved rapidly with the Compton Observatory pulsar observations.  Numerous detailed calculations have offered several possibilities for explaining the observations. For a recent summary, see \cite{Cheng}. Similarly, the discovery of gamma-ray emission from blazars stimulated an explosive growth of blazar modeling.

\item Theory gives observations more predictive power. 
For known sources, theory predicts what should be seen (or not seen) at wavelengths not yet explored. An example is the study of the region of the sky containing 3EG J1714$-$3857/SNR RXJ1713$-$3946.  CANGAROO TeV observations appeared to be consistent with a simple model of proton acceleration and interaction \cite{Enomoto}, but the theory showed that emission would then have been expected in the EGRET band.  The EGRET upper limits indicate that the simplest model is insufficient \cite{Reimer}, \cite{Butt}.

\item For the future, theory predicts what new sources should be expected. Different pulsar models, for example, make dramatically different predictions of which radio pulsars should be seen with the next generation of gamma-ray telescopes \cite{Thompson}.
\end{itemize}

\section{Conclusions}

The value of multiwavelength studies for both identification and exploration of gamma-ray sources has been clearly established. The identifications of pulsars and blazars as classes of gamma-ray emitters depended on observations at other wavelengths, and a number of other likely identifications have been directly enabled by multiwavelength observations.  The extensive theoretical work on pulsars and blazars in particular has utilized observations at many wavelengths to investigate the implications of these sources for astrophysics.

Despite substantial efforts, however, these techniques have not solved all the mysteries of the unidentified EGRET sources. A majority of the unidentified sources from the EGRET catalog remain in that category. The population studies have offered tantalizing prospects, but not definitive solutions. 

Probably the greatest advance resulting from the multiwavelength approach is the recognition of its importance and the preparation for the future.  Using a variety of methods, the multiwavelength community will be well prepared for the new generation, already in operation or planned for the near future, of satellite and ground-based gamma-ray telescopes: INTEGRAL, Swift, AGILE, GLAST, HESS, MAGIC, CANGAROO-3, VERITAS, MILAGRO.

\end{article}
\end{document}